\documentclass[pre,twocolumn,showpacs,preprintnumbers,amsmath,amssymb]{revtex4}
\usepackage{dcolumn}
\usepackage{bm}
\newcommand {\be}{\begin{eqnarray}}
\newcommand {\ee}{\end{eqnarray}}

\usepackage[pdftex]{graphicx}
\begin{document}
\title{Modeling the morphogenesis of brine channels in sea ice}

\author{B. Kutschan}
\affiliation{Ident Technologies GmbH, Rudower Chaussee 29, 12489 Berlin, Germany}

\author{K. Morawetz}
\affiliation{M\"unster University of Applied Science,
Stegerwaldstrasse 39, 48565 Steinfurt, Germany}
\affiliation{International Center for Condensed Matter Physics, Universidade de Bras\'ilia, 70904-910, Bras\'ilia-DF, Brazil}

\author{S. Gemming}
\affiliation{Forschungszentrum Dresden-Rossendorf, PF 51 01 19, 01314 Dresden, Germany}

\begin{abstract}
Brine channels are formed in sea ice under certain constraints and represent a
habitat of different microorganisms. The complex system depends on a
number of various quantities as salinity, density, pH-value or
temperature. Each quantity governs the process of brine channel
formation. There exists a strong link between bulk salinity and the presence
of brine drainage channels in growing ice with respect to both the horizontal and vertical planes. We develop a suitable phenomenological model for the
formation of brine channels both referring to the Ginzburg-Landau-theory of
phase transitions as well as to the chemical basis of morphogenesis according
to Turing. It is possible to conclude from the critical wavenumber on the size
of the structure and the critical parameters. The theoretically deduced
transition rates have the same magnitude as the experimental values. The model
creates channels of similar size as observed experimentally. An extension of
the model towards channels with different sizes is possible. The
  microstructure of ice determines the albedo feedback and plays therefore an
  important role for large-scale global circulation models (GCMs).
\end{abstract}

\date{\today}
\maketitle

\section{Introduction}

Formation and decay of complex structures depend on changes in entropy. In
 the long run structures tend to decay since the entropy of universe leads 
to a maximum and
evolves into a 'dead' steady state 
\cite{Cl}. On the other hand not only living cells
avoid the global thermodynamic equilibrium. 
{\sc A. M. Turing}\cite{Tu} showed in his paper about the chemical basis of
 morphogenesis which  additional conditions are necessary to develop
 a pattern or  structure. For instance, cells can be formed due to an
 instability of the  homogeneous equilibrium which is triggered by random
 disturbances. In this sense it should be possible,
 that the habitat of microorganisms in polar areas, the brine channels in
 sea ice, can be described through a Turing structure. 

The internal
 surface structure of ice changes dramatically when the ice cools below
 -23$^oC$ or warms above -5$^oC$ and has a crucial influence on the species
 composition and distribution within sea ice \cite{Li, KGS00}. This
 observation correlates with the change of the coverage of organisms in brine
 channels between -2$^oC$ and -6$^oC$ \cite{KGS00}.  
{\sc Golden et al}
\cite{Gol} found a critical brine volume fraction of 5 percent, or a
 temperature of -5$^oC$ for salinity of 5 parts per thousand where the ice
 distinguishes between permeable and impermeable behavior concerning 
energy and nutrient transport. According to 
{\sc Perovich et al} 
\cite{Pe} the brine volume
 increases from 2 to 37 $^o/_{oo}$  and the correlation length increases from
 0.14 to 0.22 mm if the temperature rises from  -20$^oC$ to -1$^oC$. The
 permeability varies over more than six orders of magnitude \cite{Ei}.
 Whereas 
{\sc Golden et al} 
\cite{Gol} used a  percolation model we will
 demonstrate how the brine channel distribution can  be modeled by a
 reaction-diffusion equation similar to the Ginzburg-Landau  treatment of
 phase transitions. A molecular dynamics simulation shows the change between
 the  hexagonal arranged ice structure and the more disordered liquid water
 structure \cite{Na}.


After a short introduction into the key issue of the structure formation we
describe the brine channel structure in sea ice and propose a phenomenological
description. For the interpretation of the order parameter we
discuss some microscopic properties of water using molecular dynamics
simulation in the next chapter \ref{c2}. In chapter \ref{c3} we consider the
phase transition and the conditions which allow a structure formation. We
verify the model on the basis of measured values in chapter \ref{c4} and give finally an outlook on further investigations in chapter \ref{c5}.  

\begin{figure}[h]
\centering
  \begin{center}
  \begin{tabular}{c}
  \includegraphics[width=9cm,angle=0]{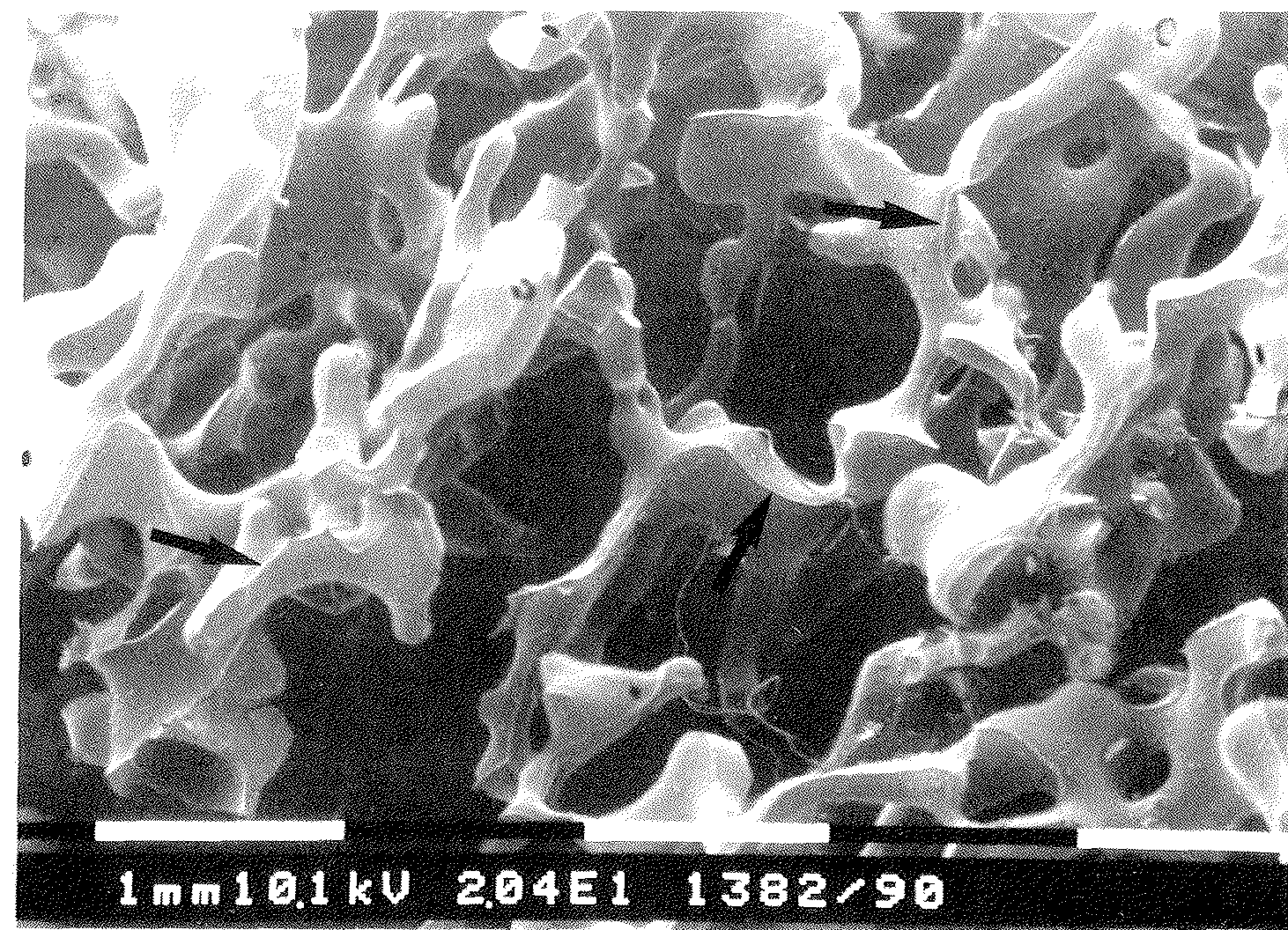}  
  \end{tabular}
  \end{center}
 \caption{\label{bb1a} SEM-image of a cast of brine channels \cite{Wei}.}
\end{figure}

\section{Microscopic properties of water \label{c2}}
\subsection{Formation of brine channels}

Various publications report on the life condition for different
groups of organisms in the polar areas in brine-filled holes, which arise under
certain boundary conditions in sea ice as base- or brine channels
(lacuna) \cite{Bar, Tr, Wei}. They are characterized by the simultaneous
existence of different phases, water and ice in a saline environment. Because
already marginal temperature variations can disturb this sensitive system,
direct measurements of the salinity, temperature, pH-value or ice crystal
are morphologically difficult \cite{Wei}. 
{\sc Weissenberger et al}
\cite{Wei1}
  developed a  cast technique in order to examine the channel
  structure. Freeze-drying eliminates the ice by sublimation, and the hardened
  casts illustrate the channels as negative pattern. Figure \ref{bb1a} shows a
  typical granular texture without prevalent orientation.  

Sometimes, both
  columnar and mixed textures occur. Using an imaging system 
{\sc Light et    al}
\cite{Li} found brine tubes, brine pockets, bubbles, drained
  inclusions, transparent areas, and poorly defined inclusions. Air
    bubbles are much larger than brine pockets. Bubbles possess a mean major
    axis length of some millimeters and brine pockets are hundred times
    smaller \cite{Pe1}. 
{\sc Cox et al} 
\cite{Co1, Co2, Co3} presented a
  quantitative model approach investigating the brine channel volume,
  salinity profile or heat expansion but without pattern formation. They
  also described the texture and genetic classification of the sea ice
  structure experimentally. A crucial factor for the brine channel structure
  formation is the spatial variability of salinity \cite{CEW99}. 

Different mechanisms are employing the mobility of brine channels
  which can be used to measure the salinity profile \cite{CEW99}. Advanced
  micro-scale photography has been developed to observe in situ the
  distribution of bottom ice algae \cite{MBMM07} which allows to determine the
  variability of the brine channel diameter from bottom to top of the ice. By
  mesocosm studies the hypothesis was established that the vertical brine
  stability is the crucial factor for ice algae growth \cite{KMG01}.
  Therefore the channel formation during solidification and its dependence on
  the salinity is of great interest both experimentally and theoretically
  \cite{WW97}. Experimentally 
{\sc Cottier et al} 
\cite{CEW99} presented
  images, which show the linkages between salinity and brine channel
  distribution in an ice sample.
 
\begin{figure}[h]
  \includegraphics[width=4.cm,angle=0]{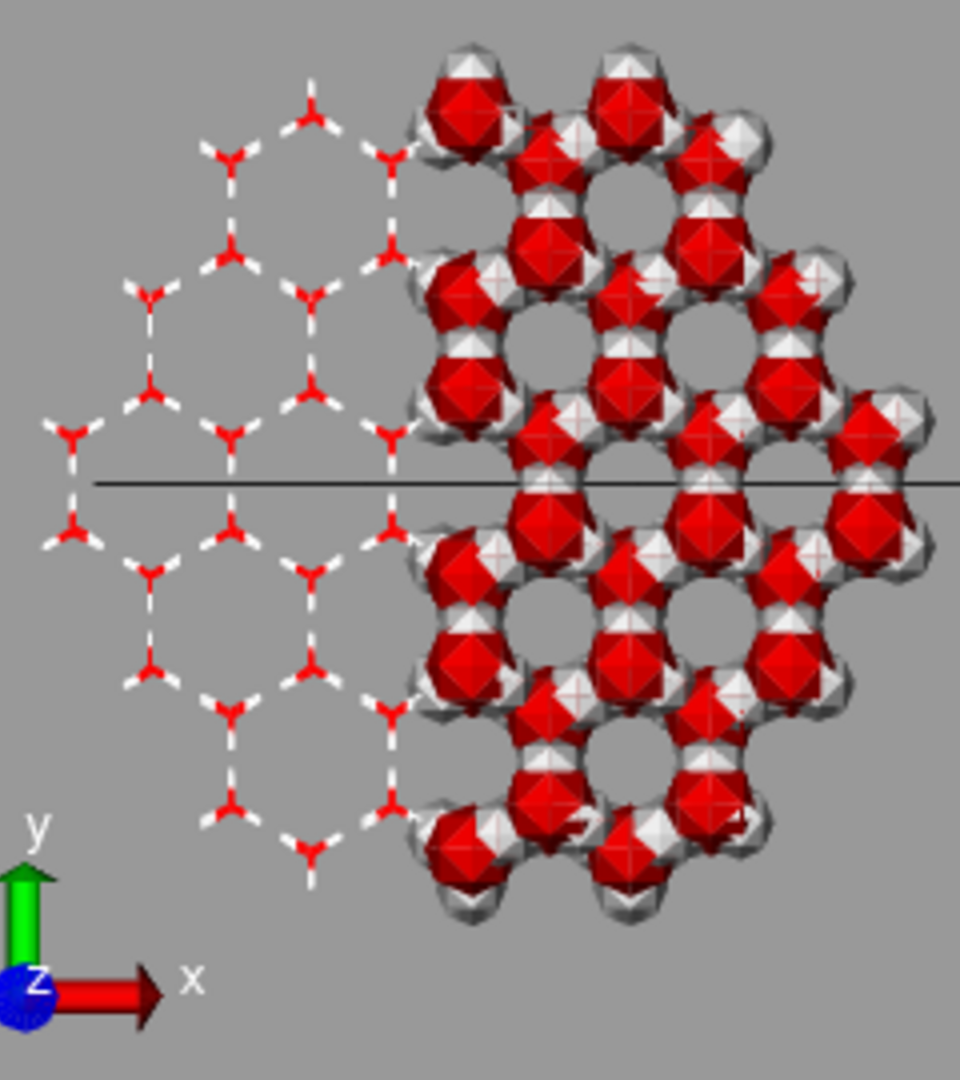}
  \includegraphics[width=3.75cm,angle=0]{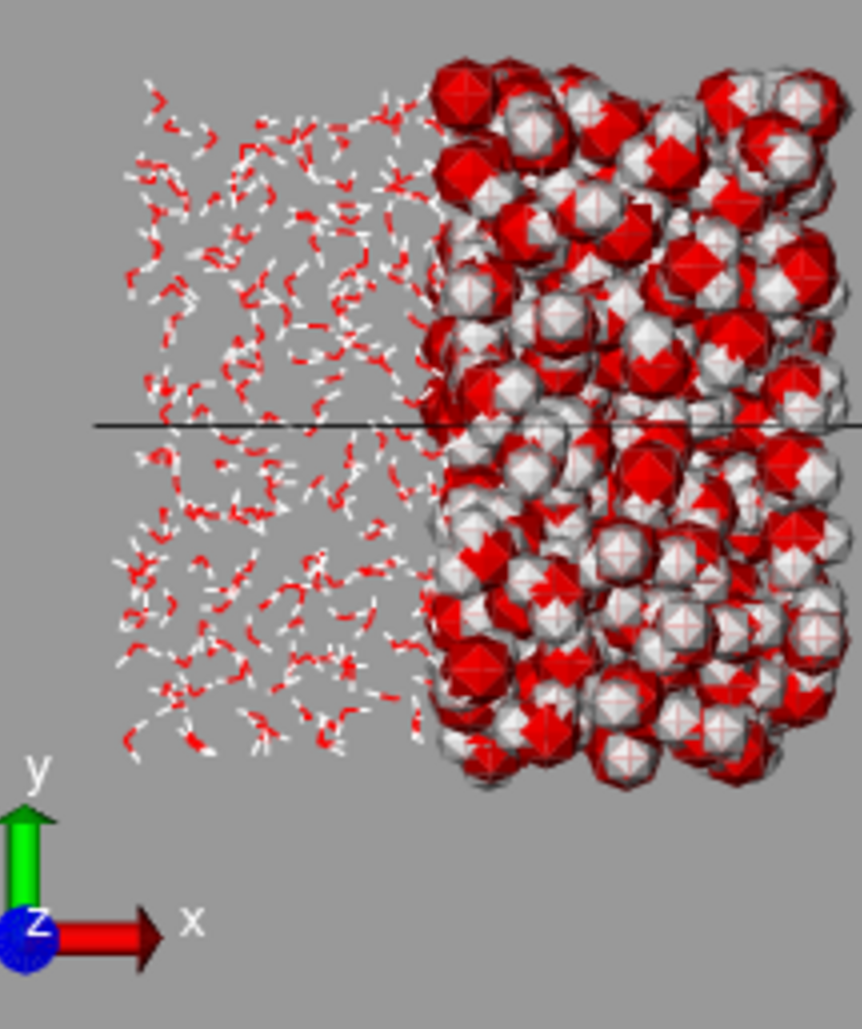}
\caption{\label{bb1b} Hexagonal ice (left) and liquid water at 300K (right).}
\end{figure}

To describe different phases in sea ice dependent on
temperature and salinity one possible approach is based on the
reaction diffusion system
\begin{equation}
\label{f1}
\boldmath \frac{\partial w(x,\mbox{\unboldmath $t$})}{\partial \mbox{\unboldmath $t$}}= f(x,\mbox{\unboldmath $t$})+D\nabla^2w(x,\mbox{\unboldmath $t$}) ,
\end{equation}

\noindent where  \boldmath  $w = \mbox{\unboldmath $\left( \begin{array}{c} u
      \\ v \end{array} \right)$}$  \unboldmath is the vector of reactants,
      \boldmath $x=\mbox{\unboldmath $ (x,y,z)$}$ \unboldmath the three
      dimensional space vector, \boldmath  $f$ \unboldmath the nonlinear
reaction kinetics and \boldmath  {\small$D = \mbox{\unboldmath $\left(    \begin{array}{cc} D_1 & 0\\0 & D_2 \end{array}  \right)$} $ \unboldmath}
the matrix of diffusitivities, where $D_1$ is the diffusion coefficient of water and $D_2$ is the diffusion coefficient of salt. For the one-
and two-dimensional case we set $y=z=0$ respectively $z=0$. The reaction
kinetics described by \boldmath $f(x,\mbox{\unboldmath $t$})$ \unboldmath can include  the theory of phase transitions by Ginzburg-Landau for the order parameters. Referring to this 
{\sc  M. Fabrizio}
\cite{Fa} presented an ice-water phase transition. Under certain
conditions, spatial patterns evolve in the so-called Turing space. These
patterns can reproduce the distribution ranging from sea water with high salinity  to sea
ice with low salinity. The brine channel system exists below a critical
temperature in a thermodynamical  non-equilibrium. It is driven via the
desalination of ice during the freezing process that leads to a salinity
increase  in the brine channels. The higher salt concentration in the
remaining liquid phase leads to a freezing point depression and triggers the
ocean currents.

\subsection{Different states of water}

Already {\sc Wilhelm Conrad R\"ontgen}  had described the anomalous properties of
water with molecules of the first kind, which he called ice molecules and
molecules of the second kind \cite{Roe} and which represent the liquid aggregate state. 
{\sc  Dennison}
\cite{De}  
determined the ordinary hexagonal ice-I modification from X-ray pattern methodically verified by 
{\sc Bragg}
\cite{Br}. This so called $E_h$-ice is formed by four oxygen atoms which
build a tetrahedron as illustrated in figure \ref{bb1b}. In $E_h$-ice each
oxygen atom is tetrahedrically coordinated by four neighbouring oxygen atoms,
each accompagnied by a hydrogen bridge.  
The arrangement is isomorphous to the wurtzite form of
zinc sulphide or to the silicon atoms in the tridymite form of silicon
dioxide. 
{\sc Bjerrum} 
\cite{Bj} and 
{\sc Eisenberg} 
\cite{Eis} have provided a
survey about the structure differences between the different polymorphic forms
of ice and liquid water.

Molecular dynamics simulations with the TIP3P-model of water using the
NAMD-software by the Theoretical and Computational Biophysics Group of the
university of Illinois show the change from a regular hexagonal lattice
structure to irregular bonds after the melting (figure \ref{f1a}). 
{\sc Nada  et al} 
\cite{Na} developed a better six-site potential model of $H_2O$ for a crystal
growth of ice from water using molecular dynamics and Monte Carlo
methods. They computed both the free energy and an order parameter for the
description of the water structure. Also 
{\sc Medvedev et al} 
\cite{Me} introduced a ''tetrahedricity measure'' $M_T$
for the ordering degree of water. It is possible to discriminate between
ice- and water molecules via a two-state function ($G=0$ if $M_T \ge M^c_T$ and
$G=1$ if  $M_T < M^c_T$). This tetrahedricity is computed using the sum
\begin{equation}
\label{f1a}
M_T=\frac{1}{15 <l^2>}\sum_{i,j}(l_i - l_j)^2 ,
\end{equation}  
where $l_i$ are the lengths of the six edges of the tetrahedron
formed by the four nearest neighbors of the considered water molecule. For an
ideal tetrahedron one has $M_T=0$ and the random structure yields $M_T=1$. The
tetrahedricity can be used in order to define an order parameter according to
the Landau - de Gennes model for liquid crystals, which refers to the
Clausius-Mosotti-relation. Other simulations such as the percolation models of 
{\sc Stanley et al} 
\cite{St1, St2} use a two-state model, in which a critical correlation length determines the phase
transition. A mesoscopic model for the sea ice crystal growth is
  developed by 
{\sc Kawano} and {\sc Ohashi} 
\cite{Ka} who used a Voronoi  dynamics. 

\section{Reaction - diffusion model  \label{c3}}
\subsection{1+1-dimensional model equations}

We consider the reaction diffusion system
\begin{eqnarray}
  \frac{\partial u(x,t)}{\partial t} & = & a_1 u - c u^3 +d u^5 + b_1 v + D_1
  \frac{\partial^2 u(x,t)}{\partial x^2} \label{f2}\\
  \frac{\partial v(x,t)}{\partial t} & = & -a_2 v -  b_2 u + D_2 
  \frac{\partial^2 v(x,t)}{\partial x^2} \label{f3}
\end{eqnarray}
in one space dimension. The order parameter according to the Ginzburg-Landau-theory is $u(x,t)$ with
$u_{min} \leq u_c \leq u_{max}$ and proportional to the tetrahedricity
$u \sim M_T$. If the variable $u$ is smaller than $u_c$ ($u < u_c$),
the phase changes from water to ice and vice versa. Thus, changes in $u$ reflect
 temperature variations. The variable $v$ is a
measure of the salinity. The coefficient $a_1$ depends on the temperature $T$
as $(T-T_c)/T_c$ with the critical temperature $T_c$. The salt
exchange between ice and water is realized by the gain term $b_1v$ and the
loss term $-a_2 v$. The positive terms  $a_1u$ and $b_1v$ are the temperature-
and salt-concentration-dependent ''driving forces'' of the system. 

\begin{figure}[h]
\centering
  \includegraphics[width=9cm,angle=0]{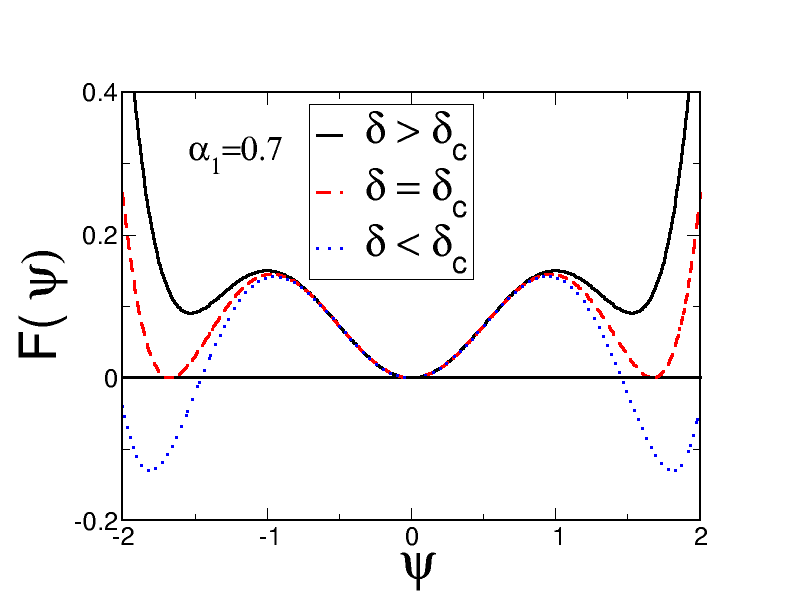}
 \caption{\label{bbb1} Landau-function (\ref{lf1}) versus the dimensionless order parameter (tetrahedricity) for various $\delta$.}
\end{figure}

In order to
realize the $T$-dependent phase transition, one can expand the order parameter in a power series corresponding to
Ginzburg and Landau in equation \ref{f2}. In order to describe properly a temperature-induced phase transition of second order an expression $-c u^3$ is necessary. The first-order phase transition is dependent on $d u^5$. Supercooled or superheated phases can coexist, i.e. an hysteresis
behavior is possible.  Without the term $d u^5$ we can also realize a
  brine channel formation. But this second-order phase transition does
  not allow us to consider the specific heat as a jump in the 
order parameter $\psi$ (figures \ref{bbb1} and \ref{bbb2}). We write the 
equations (\ref{f2}) and
(\ref{f3}) in dimensionless form (see appendix) by setting  $\tau=\sqrt{b_1b_2}t$,
$\xi=\sqrt[4]{{b_1 b_2}/{D^2_1}}x$, $\psi=\sqrt[4]{{c^2}/{b_1 b_2}}u$,
$\rho=\sqrt[4]{{b_1 c^2}/{b^3_2}}v$ and get 
\begin{eqnarray}
  \frac{\partial \psi(\xi,\tau)}{\partial \tau} & = &
  f[\psi,\rho] +
  \frac{\partial^2\psi(\xi,\tau)}{\partial \xi^2}
  \label{f4}\\
  \frac{\partial \rho(\xi,\tau)}{\partial \tau} & = &
  g[\psi,\rho] + D \frac{\partial^2
  \rho(\xi,\tau)}{\partial \xi^2} \label{f5} .
\end{eqnarray}
with $\alpha_1={a_1}/{\sqrt{b_1 b_2}}$, $\delta={d
  \sqrt{b_1 b_2}}/{c^2}$, $\alpha_2={a_2}/{\sqrt{b_1 b_2}}$ and
  $D={D_2}/{D_1}$ as well as the reaction kinetics 
\begin{eqnarray}
 f[\psi,\rho] & = & \alpha_1 \psi - \psi^3 +\delta\psi^5
 +\rho  \label{f5a} \\  
 g[\psi,\rho] & = & -\alpha_2 \rho - \psi,  \label{f5aa}
\end{eqnarray} 
where $\psi$ is the dimensionless order parameter of the water/ice system and $\rho$ the dimensionless salinity. Thus the dynamics only depends on  four
 parameters  $\alpha_1$, $\alpha_2$, $\delta$ and $D$. Without the salinity
 $\rho$ in equation (\ref{f2}) respectively (\ref{f4}) the
 above equation system is reduced to a Ginzburg-Landau equation for 
the first order phase transition.

\subsubsection{First Order Phase Transitions\label{ue1}}

When we neglect the salinity $\rho$, the integration of the kinetic
function (\ref{f5a}) yields the Landau function for the order parameter $\psi$
of water/ice 
\begin{eqnarray}
 F &=&  \frac{a_1}{2} \psi^2 - \frac{1}{4}\psi^4 + \frac{\delta}{6}
 \psi^6 
\label{lf1} \end{eqnarray} 
as plotted in figure~\ref{bbb1}. It possesses three minima
\begin{equation}
\psi_{\rm min} =
\left\{0,\pm\sqrt{\frac{1}{2\delta}\left(1+\sqrt{1-4\delta\alpha_1}\right)}\right\} .
\label{lf2}
\end{equation}
When several different
minima of equal depth exist, then there is a discontinuity in $\psi$
due to Maxwell construction and one has a first-order phase transition
\cite{Hau}. This is the case if

\begin{equation}
\delta=\delta_c=\frac{3}{16\alpha_{1c}} \label{5b}
\end{equation}
and 
\begin{equation} 
\psi^2_c=\frac{3}{4\delta_c} ,
\end{equation}
in consequence of
\begin{equation}
F=\psi^2\left(\frac{\delta}{6}\left(\psi^2-\frac{3}{4\delta}\right)^2+\frac{\alpha_1}{2}-\frac{3}{2\cdot
  16 \delta}\right) = 0 .
\end{equation}

\begin{figure}[h]
\centering
  \includegraphics[width=9cm,angle=0]{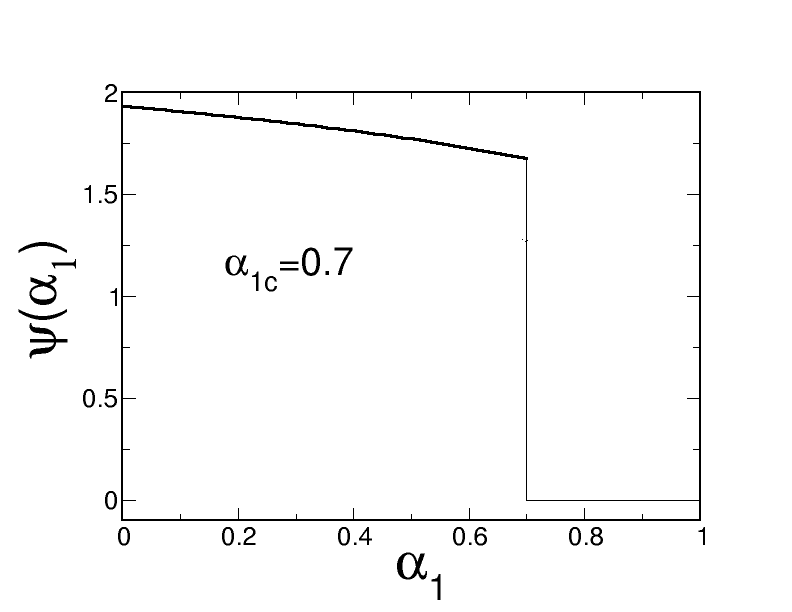}
\caption{\label{bbb2} 
The minimal order parameter
  $\psi_{\rm min}$ of (\ref{lf2}) dependent on $\alpha_1$.}
\end{figure}

Thus the critical parameter $\delta=\delta_c$ is determined by the
temperature-dependent critical value $\alpha_1=\alpha_{1c}$. The jump in 
figures \ref{bbb1} and \ref{bbb2} is a measure for the latent heat of the phase
transition from water to ice.  
{\sc Feistel} and {\sc Hagen} 
\cite{Fei}
 have deduced theoretically the latent heat of sea ice for various salinities.

\subsection{Linear Stability Analysis}

First we perform a linear stability analysis by linearizing the equation
 system (\ref{f4}) and (\ref{f5}) according to $\psi=\psi_0 + \bar \psi$ and
 $\rho=\rho_0 + \bar \rho$. We obtain the characteristic equation for
the fixed points with 
\be
\left (\begin{array}{l}
\bar \psi\\\bar \rho 
\end{array}\right ) 
=\left ( \begin{array}{l}
\bar \psi_i\\\bar \rho_i 
\end{array}\right ) 
exp(\lambda(\kappa) \tau
+i\kappa\xi)
\ee
as
\begin{eqnarray}
 \left( \begin{array}{c}
    \frac{\partial}{\partial{\tau}}\bar\psi\\\frac{\partial}{\partial{\tau}}\bar\rho  \end{array}
\right) &=&   \left( \begin{array}{cc}\alpha_1-3\psi^2_0 +5\delta\psi^4_0 & 1\\-1 & -\alpha_2 \end{array}
\right) \left( \begin{array}{c} \bar\psi\\\bar\rho  \end{array} \right) \nonumber\\ &+& \left(
  \begin{array}{cc} \frac{\partial^2}{\partial{\xi^2}} & 0\\0 & D \frac{\partial^2}{\partial{\xi^2}} \end{array}
\right) \left( \begin{array}{c} \bar\psi\\\bar\rho  \end{array} \right)   \label{f7}
\end{eqnarray}
as outlined in appendix.

There are five fixed points for the kinetics (\ref{f5a}) and
(\ref{f5aa}) which satisfy $f=0$ and $g=0$. In order to get a stable
non-oscillating pattern we need a stable spiral point as fixed point. Moreover
the associated eigenvalues have to possess a positive real part for a positive
wavenumber, i.e. they have to allow to create unstable modes. 
Not each fixed point satisfies both conditions. Therefore for
the following discussion we choose the steady state, $\psi_0=0$ and
$\rho_0=0$, which corresponds to the observable brine channel structures
measured by a casting experiment \cite{MBMM07,CEW99}.
\begin{figure}[h]
\centering
 \includegraphics[width=9cm,angle=0]{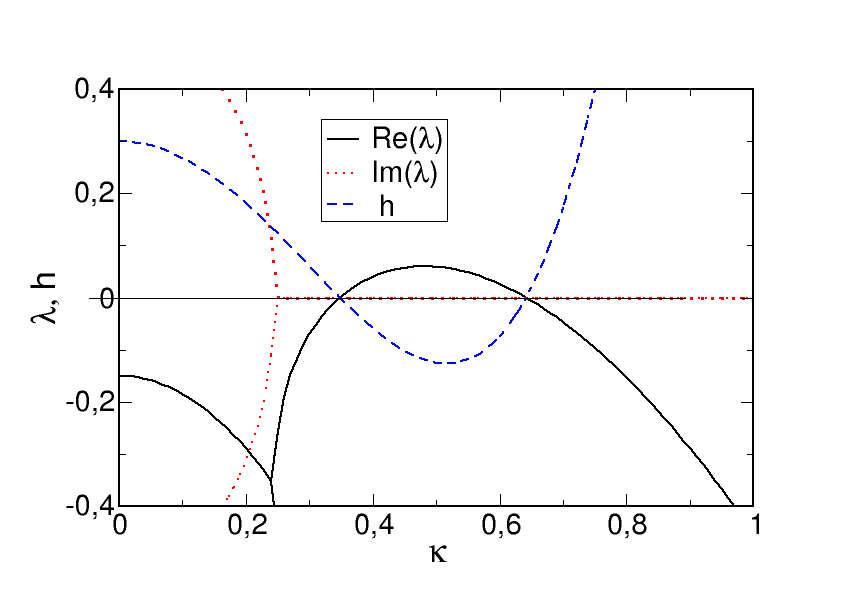}  
 \caption{\label{figure5} Dispersion of the linear stability (\ref{sol}) versus the dimensionless wave number $\kappa$ for
   $\alpha_1=0.7$, $\alpha_2=1$, $D=6$ together with the function $h(\kappa)$
 of (\ref{14})}
\end{figure}

 Short-time experiments
may also record structures that are formed under non-steady conditions. Since
those structures are beyond the scope of the present paper, we proceed with
the steady state which leads from (\ref{f7}) to
\begin{equation}
 \lambda(\kappa)^2 +[\kappa^2(1+D)+\alpha_2-\alpha_1]\lambda(\kappa)+h(\kappa^2)=0
\label{lam}
\end{equation}
with
\begin{equation}
 h(\kappa^2)=D\kappa^4+(\alpha_2-\alpha_1 D)\kappa^2-\alpha_1 \alpha_2+1
\label{14}
\end{equation}
and which is readily solved
\be
\lambda(\kappa)_{1,2}&=&\frac 1 2 \left (\alpha_1 - \alpha_2 - (1 + D) \kappa^2 
\right . \nonumber\\&&\left .
\pm\sqrt {(\alpha_1+\alpha_2+(D-1)\kappa^2)^2-4} \right ).
\label{sol}
\ee

\subsection{Turing space}

Let us discuss the equations (\ref{lam}) and (\ref{14}) concerning the conditions for the occurrance of Turing structures in detail. First we concentrate on the situation $\kappa=0$ where a homogeneous phase is formed. 
Then the fixed points $\psi=0$ and $\rho=0$ are stable according to the eigenvalues (\ref{sol}) if
\begin{eqnarray}
 \lambda_{1,2}(0)&=&-\frac{\alpha_2-\alpha_1}{2}\pm\frac{1}{2}\sqrt{(\alpha_1-\alpha_2)^2-4 (1-\alpha_1\alpha_2)}\nonumber\\&&
\label{hom}
\end{eqnarray}
are negative. Otherwise we would have a globally unstable situation
which we rule out. Also homogeneously oscillations don't describe a
brine channel formation. It is easy to see that the solution (\ref{hom}) gives only two negative values if 
\be
{\rm condition}\, {\rm I}:\quad \alpha_2>\alpha_1\,\, {\rm and}\,\, \alpha_1 \alpha_2<1.
\label{cond1}
\ee
The trajectories
of salinity $\rho$ and the order parameter $\psi$ converge to
the steady state value zero by damped oscillations. Therefore the structure formation does not follow from the
initial condition in time but from the range of the interaction in space.  

Next we discuss the spatial inhomogeneous case, $\kappa^2>0$, where some spatial fluctuations may be amplified and form macroscopic structures, i.e. the Turing structure. Therefore we search for such modes of (\ref{sol}) which grow in time, i.e. ${\rm Re}\lambda(\kappa)>0$. Time oscillating structures appear if ${\rm Im} \lambda(\kappa)\neq 0$ which can be seen from the solution of (\ref{sol}) to be the case if $\alpha_1+\alpha_2-2<(1-D)\kappa^2<\alpha_1+\alpha_2+2$. In this region we have
\be
{\rm Re} \lambda_{\rm osc}(\kappa)={\alpha_1-\alpha_2-\kappa^2 (1+D)\over 2}.
\ee
Demanding to be positive means $\kappa^2 (1+D)<\alpha_1-\alpha_2$. Due to
(\ref{cond1}) this cannot be fulfilled
since the diffusion constant $D$ is positive and $\kappa$ real. Therefore for
a time-growing mode we do not have an imaginary part of $\lambda$ in our
model. In other words we do not have oscillating and time-growing structures.
The restriction for the only allowed region is
\be
|(1-D)\kappa^2-\alpha_1-\alpha_2|>2.
\label{cond1a}
\ee
In this region we search now for the condition $\lambda>0$. The term before the square in (\ref{sol}) is negative as can be seen from (\ref{cond1}). Therefore we can only have positive $\lambda$ if the square of this term is less than the content of the root. This leads to 
\be
(\alpha_1-\kappa^2)(\alpha_2+D\kappa^2)>1
\ee
which restricts the $\kappa$ region to the interval
\be
\kappa^2\in{1\over 2D} \left (\alpha_1 D-\alpha_2\pm\sqrt{(\alpha_1 D+\alpha_2)^2-4 D} \right).
\label{cond1b}
\ee
Due to (\ref{cond1})  the term under the square root is samller than the square of the first term in (\ref{cond1b}) and we get only a meaningful condition from (\ref{cond1b}) if
\be
\alpha_1 D>\alpha_2.
\label{cond2a}
\ee
Moreover, the square root must be real, i. e.
\be
(\alpha_1 D+\alpha_2)^2-4 D > 0,
\label{cond2b}
\ee
which leads to 
\be
&&D<{(1-\sqrt{1-\alpha_1\alpha_2})^2\over \alpha_1^2}\quad {\rm or}\quad D>{(1+\sqrt{1-\alpha_1\alpha_2})^2\over \alpha_1^2}.\nonumber\\&&
\ee
This has to be in agreement with (\ref{cond2a}) and
discussing the different cases results finally into 
\be
{\rm condition}\, {\rm II}:\qquad D>{(1+\sqrt{1-\alpha_1\alpha_2})^2\over \alpha_1^2}.
\label{cond2}
\ee

Having determined the ranges of $\alpha_1$, $\alpha_2$ and $D$
we have to inspect the two conditions on $\kappa$, i.e. (\ref{cond1b}) and
(\ref{cond1a}). Discussing separately the cases $D\gtrless 1$ one sees that
(\ref{cond1a}) gives no restriction on (\ref{cond1b}). 

\begin{figure}
  \includegraphics[width=9cm,angle=0]{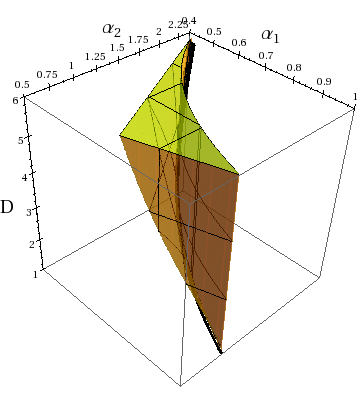}
\caption{The Turing space as phase diagram where spatial structures can
  occur. The lower limiting line, $\alpha_2=1/\alpha_1$, $D=1/\alpha_1^2$, is plotted as thick line.
\label{turinga}}
 \end{figure}

\begin{figure}
  \includegraphics[width=9cm,angle=0]{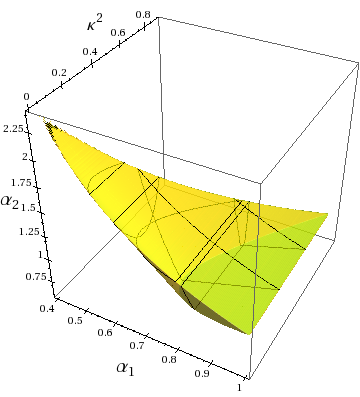}
\caption{The possible wave lengths $\kappa^2$ where spatial structures can
  occur for $D=6$ in dependence on $\alpha_1$ and $\alpha_2$.
\label{turing1a}}
 \end{figure}

Collecting now all conditions for the occurrance of a Turing structure,
(\ref{cond2}), (\ref{cond1}) and (\ref{cond1b}), we obtain
\be
{\rm cond.}\, {\rm I}:&&\quad \alpha_2\ge\alpha_1\,\, {\rm and}\,\, \alpha_1 \alpha_2\le1.
\nonumber\\
{\rm cond.}\, {\rm II}:&&\quad D\ge{(1+\sqrt{1-\alpha_1\alpha_2})^2\over \alpha_1^2}
\nonumber\\
{\rm cond.}\, {\rm III}:&&\kappa^2\in{1\over 2D} \left (\alpha_1
  D-\alpha_2\pm\sqrt{(\alpha_1 D+\alpha_2)^2-4 D} \right).
\nonumber\\
&&
\label{cond}
\ee
The Turing space as phase diagram is determined by condition I and II and is 
plotted in figure \ref{turinga}.
One can see that the Turing space starts at the minimal (tricritical point)
\be
\alpha_{1t}=\alpha_{2t}=D_t=1
\ee
which means that we have only a Turing space for sufficient large diffusivity
$D\ge 1$. For the Turing space we obtain the possible wave numbers according
to condition III as plotted in figure \ref{turing1a}. 

\begin{figure}[h]
\includegraphics[width=9cm,angle=0]{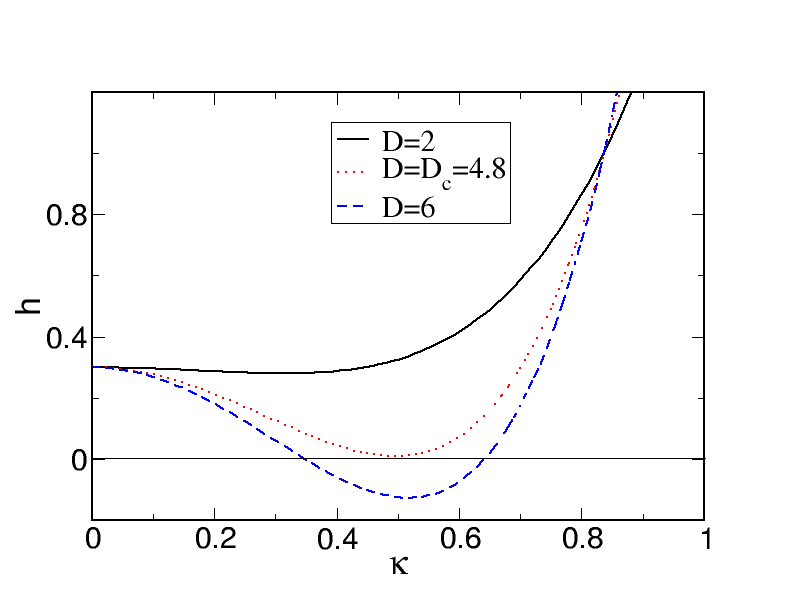}
\caption{\label{h} Dispersion $h(\kappa)$ for different 
$D$, $\alpha_1=0.7$ and $\alpha_2=1$. }
\end{figure}

\subsection{Critical modes}

The critical wavenumber can be found from the largest modes. These are given
by the minimum of equation (\ref{14}) from which we
find the wavenumbers
 \begin{equation}
 \kappa^2_{min}=\frac{1}{2D}(Df_\psi+g_\rho)=\frac{1}{2D}(D\alpha_1-\alpha_2) \label{kmin}
\end{equation}
and the minima
\begin{equation}
 h_{min}=f_\psi g_\rho-g_\psi
 f_\rho-\frac{(Df_\psi+g_\rho)^2}{4D}=1-\frac{(D
   \alpha_1+\alpha_2)^2}{4D}.  \label{hmin}
\end{equation}

\begin{figure}[h]
  \includegraphics[width=9cm,angle=0]{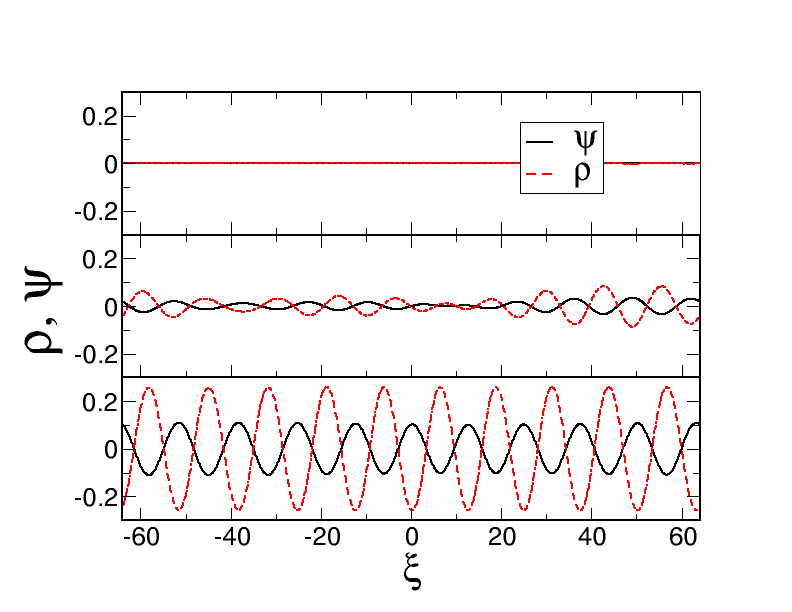}
   \caption{\label{b3} Time evolution of the order parameter
   $\psi$ and salinity $\rho$ versus spatial coordinates for    
$\tau=100, 170, 400$ (from above to below)for $\alpha_1=0.7$, $\alpha_2=1$,
  $\delta=\frac{3}{16\alpha_1}$, $D=6$ with the initial condition
  $\rho(\tau=0)=0.5 \pm 0.01N(0,1)$ and periodic boundary conditions.}
\end{figure}

\noindent For $\kappa^2_{min}>0$ and $h_{min} < 0$ we find again the
corresponding inequalities (\ref{cond2a}) and (\ref{cond2b}). The
formation of a spatial Turing structure, a non-oscillating pattern,
requires a negative $h_{min}$ for $\kappa^2_{min}>0$. In this case
there is a range of wavenumbers which are linearly unstable as seen in figure 
\ref{figure5}. In figure \ref{h} we illustrate the behaviour of $h(\kappa)$
for different diffusion constants. Only those which lead to negative $h$
are forming the Turing structure as discussed in the previous chapter.
This critical range
can be obtained, if the diffusion coefficient $D$ is greater than the critical
diffusion coefficient of condition II, (\ref{cond}) $D_c$
\begin{equation}
 D_c={(1+\sqrt{1-\alpha_1\alpha_2})^2\over \alpha_1^2},
\end{equation}
which we get from $h_{min}=0$ with the critical wavenumber $\kappa_c$
\begin{equation}
 \kappa^2_c=\frac{D_c f_\psi+g_\rho}{2D_c}=\frac{D_c \alpha_1-\alpha_2}{2D_c} .\label{fkc}
\end{equation}

\begin{figure}[h]
\centering
 \includegraphics[width=9cm,scale=0.9,angle=0]{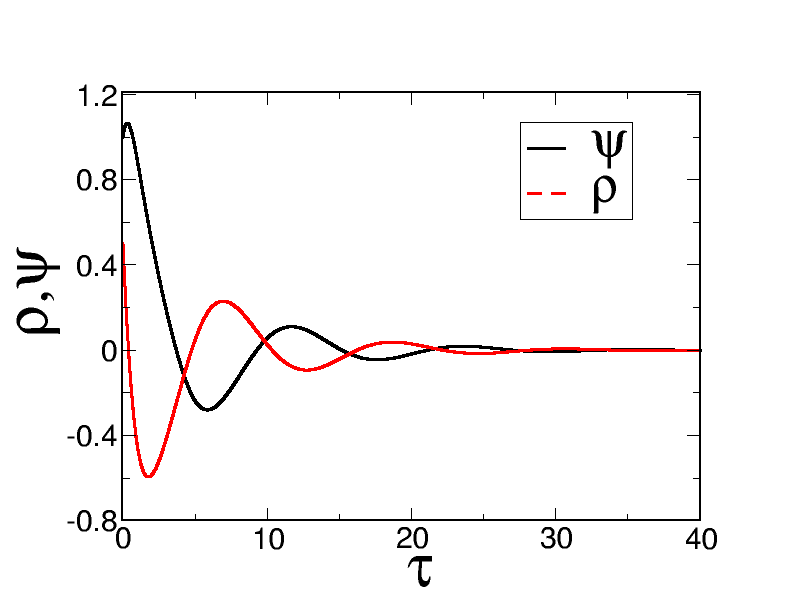}
 \caption{\label{kin} Time evolution of the order parameter $\psi$ and the salinity $\rho$ for $\alpha_1=0.7$, $\alpha_2=1$,
 $\delta=\frac{3}{16\alpha_1}$ and the initial order parameter $\psi(\tau=0)=1$ and the dimensionsless salinity $\rho(\tau=0)=0.5$.}
\label{timeev}
\end{figure}

The size of the structure can be estimated from
$\frac{2\pi}{\kappa_c}$. The pattern size depends on the both
parameters $\alpha_1$ and $\alpha_2$. The parameters
determine the brine channel size and vice versa. With the parameters
chosen in fig. \ref{figure5} we obtain 
a pattern size of $12.6$. In the next chapter we
compare this value with experimental quantities. 
With a small initial random perturbation we plot snapshots of
the time evolution of the order parameter $\psi$ and the salinity $\rho$ in 
figure \ref{b3}.  The quantities
$\psi$ and $\rho$ are opposite to each other; domains with low salinity correspond to domains
with ice and domains with high salinity to water domains.  We see the formation of a mean mode given by the wave length
$\kappa_c$.

A positive $h(\kappa=0)$, respectivly a negative $\lambda(\kappa=0)$, for
$\kappa=0$ guarantees that $\rho$ and $\psi$ converge to the stable fixed
point $\rho_0=0,\psi_0=0$. Therefore the structure formation
does not follow from the initial oscillations in time
(fig. \ref{timeev}). In order to obtain a new spatial structure there must
exist at least a negative $h$ for $\kappa>0$, respectively a positive
eigenvalue $\lambda$, for $\kappa>0$ as was discussed in the last chapter.

\subsection{2+1-dimensional model}

\begin{figure}[htb!]
\centering
\includegraphics[width=9cm,scale=1.0,angle=0]{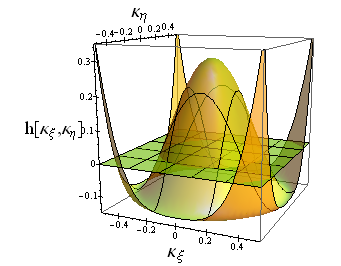}
 \caption{\label{b4} Dispersion of the two-dimensional characteristics (\ref{20}) and (\ref{21}) for $\alpha_1=0.7$, $\alpha_2=1$, $D=6$.}
\end{figure}

\noindent From the characteristic equation in the spatially two-dimensional case
  \begin{equation}
 \lambda^2 +[(\kappa^2_\xi+\kappa^2_\eta)(1+D)+\alpha_2-\alpha_1]\lambda+h(\kappa^2_\xi,k^2_\eta)=0
\label{20}
\end{equation}
we find the corresponding dispersion relation 
\begin{equation}
 h(\kappa^2_\xi,\kappa^2_\eta)=D(\kappa^2_\xi+\kappa^2_\eta)^2+(\alpha_2-\alpha_1D)(\kappa^2_\xi+\kappa^2_\eta)-\alpha_1\alpha_2+1 
\label{21}
\end{equation}
which is illustrated in figure \ref{b4}.

\begin{figure}[h]
  \includegraphics[width=9cm,angle=0]{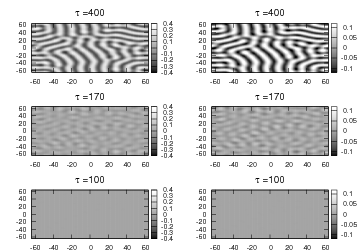}
   \caption{\label{b5} Structure formation for 3 time steps $\tau=100,170,400$ (from top to bottom) for the order parameter $\Psi$ (left) and the salinity $\rho$ (right). The parameters are $a_1=0.7$, $a_2=1$,
  $d=\frac{3}{16a_1}$, $D=6$ with the initial condition
  $v(t=0)=0.5 \pm 0.01N(0,1)$ and periodic boundary conditions.}
\end{figure}

\begin{figure}[h]
  \includegraphics[width=9cm,angle=0]{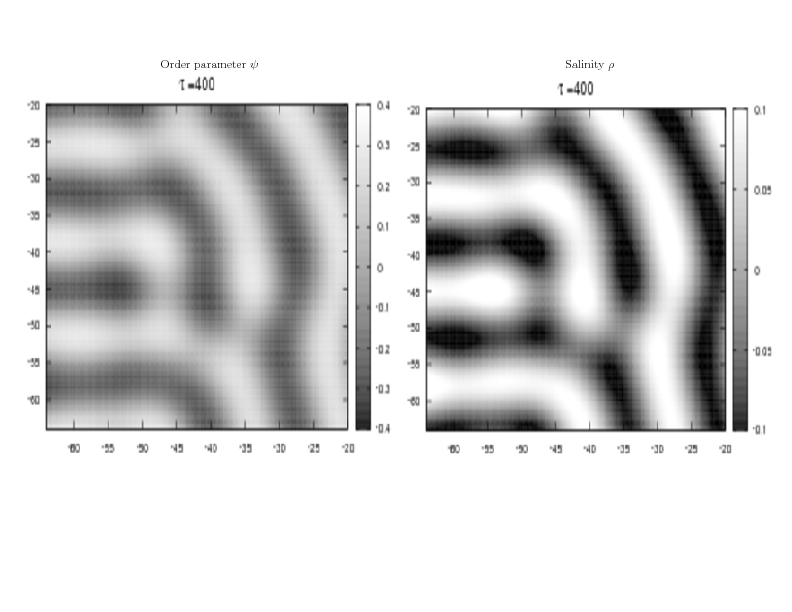}
   \caption{\label{b5a}  
   Magnification of a detail of figure~\ref{b5}.}
\end{figure}

The Turing space is bounded by the sectional plane $h=0$. The
evolution of the order parameter $\psi$ and the salinity $\rho$ is illustrated
in figures \ref{b5} and \ref{b5a}. Their behavior is inversely proportional
and corresponds to the fact, that a high salinity occurs in the water phase 
and a low
salinity in the ice phase. Similiar as in the one-dimensional case we see the dominant formaton of one wavelength. The model kinetics generate brine channels of similar
size. In order to obtain a hierarchical net of brine channels of different
size, the kinetics in the basic equations can be altered accordingly \cite{Mu}.

\subsection{Note on second and thirth order kinetics}

If we replace the kinetics (\ref{f5a}) by 
\begin{equation}
 f[\psi,\rho]=\alpha_1\psi-\psi^3+\rho
\label{}
\end{equation}
or
\begin{equation}
 f[\psi,\rho]=\alpha_1\psi-\psi^2+\rho
\label{}
\end{equation}
we can carry out the same linear stability analysis for the fixed point
$\psi_0=0$ and $\rho_0=0$. Then we obtain the same characteristic
equation as (\ref{lam}) and (\ref{14}). Therfore we get the same
Turing space for the structure formation. Consequently, 
both kinetics allow us to
realize a brine channel formation but the thirth order kinetics
describes a second order phase transition only. In this connection it
is possible to discuss second order phase transitions with
spin models, too.  

\section{Connection to experimental data \label{c4}}

The critical domain size is determined by the equation (\ref{fkc}). Due to
this relation we can infer other parameters in the model
equations  (\ref{f2}) and (\ref{f3}). From the relation between the
dimensionless wavenumber $\kappa$ and the dimensional wavenumber $k$ 
\begin{equation}
 \kappa^2=\frac{\alpha_1 D-\alpha_2}{2D}=\frac{D_1}{\sqrt{b_1b_2}}k^2 \label{f11}
\end{equation}
we get
\begin{equation}
 \frac{2 \pi}{\kappa_c}=12.6=\frac{2 \pi}{k_c}
 \frac{\sqrt[4]{b_1b_2}}{\sqrt{D_1}}  .
\end{equation} 

The observed diameters of the brine channels range from  $\mu
 m$ to $mm$ scale \cite{Wei}. For a size of ${2 \pi}/{k_c}=10 \mu
m$ and a diffusion coefficient $D_1=10^{-5} {cm^2}{s^{-1}}$  for
$H_2O$-molecules we obtain the product $b_1b_2=2.5 \cdot 10^6 s^{-2}$ and
a transition rate $a_1=\sqrt{b_1b_2}\alpha_1=1111 s^{-1}$. The rate $a_1$ is
proportional to reorientations of the molecules per second,
${1}/{\tau_d}=10^5 {s^{-1}}$ 
({\sc Eisenberg})
\cite{Eis} and to the scaled temperature
$\frac{T_c-T}{T_c}$

\begin{equation}
a_1 \sim \frac{T_c-T}{T_c}\frac{1}{\tau_d}  , \label{f12}
\end{equation}

\noindent where $T_c$ is the melting point depending on the salinity. The mean
salinity in sea ice of $35 {g}/{l}$ corresponds to 1 $NaCl$-molecule per
100 $H_2O$-molecules, i.e. 1 $Na^+$-ion and 1 $Cl^-$-ion per 100
$H_2O$-molecules in a diluted solution after the dissociation or a ratio of
$x=({n_{Na^+}+n_{Cl^-}})/{n_{H_2O}}={1}/{50}$. From this facts we
obtain according to Clausius-Clapeyron 
\begin{equation}
 \Delta T = -\frac{xRT^2}{\Delta H} 
\end{equation} 
a freezing point depression from $0 ^\circ C$ to $-2 ^\circ C$,
where $\Delta H=6\frac{kJ}{mol}$ is the latent heat of the phase transition
from water to ice, $R=8.314 \frac{J}{mol K}$ is the universal gas constant and
$T=273 K$. Thus we obtain correctly $T_c=271 K$. For an environmental temperature
of $T=-5
^\circ C=268 K$  according to (\ref{f12}), a transition rate of $\frac{271-268}{271}\cdot
10^5 s^{-1}=1107 s^{-1}$ follows which nearly corresponds to $a_1=1111 s^{-1}$, which we
estimated from the domain size (\ref{f11}). 

Furthermore we find the transition
rate $a_2=\sqrt{b_1 b_2}\alpha_2=1587 s^{-1}$ and the diffusion coefficient
$D_2=6 \times 10^{-5}\frac{cm^2}{s}$. Due to the transformation of
  equation (\ref{f2}) into the dimensionless form (\ref{f4}) there  exists
a fixed relation (\ref{5b}) because of $c=1$. We obtain for the equation
(\ref{f2}) the rate $d_c$  with $\alpha_1={a_1}/{\sqrt{b_1 b_2}}$, 
$\delta={d \sqrt{b_1 b_2}}/{c^2}$ of
\begin{equation}
 d_c = \frac{3 c^2}{16 a_1} , 
\end{equation}
which is proportional to $c^2$. A transition rate $c=1000s^{-1}$ yields
a critical rate $d_c=169s^{-1}$. From the knowledge of the diffusion
coefficient $D_1$ and the size of brine channels we can deduce
the two rates $a_1$ and $a_2$. Both rates possess the same order of
magnitude and are inside the Touring space of structure formation. 
If the experiments would lead to other parameters $a_1$ and $a_2$, 
especially to rather different values, a brine channel could not arise because 
of the limitation of the Turing space in figure \ref{turinga}
and \ref{turing1a}. In other words the model here seems to describe the experimental 
finding of brine channel formation. 

Due to the small difference between the time
constants $a_1$ and $a_2$ we obtain a dynamic interference between the
reorientation of the water molecules and the desalinization. Both are evolving on nearly the same time scale. 
In particular, we
cannot simplify the kinetics by separating time scales using the Tichonov theorem \cite{Ti1952} in order to reduce our reaction
diffusion system (\ref{f2}),(\ref{f3}) but have to consider both dynamics as demonstrated.

\section{Conclusion  \label{c5}}

In this paper it has been shown that a  reaction diffusion
system which connects the basic ideas both of Ginzburg and of Turing can
describe the formation of brine channels with realistic parameters. 
For the chosen parameters patterns of similar
size emerged. 
{\sc Eicken} 
\cite{Ei1} and 
{\sc Weissenberger} 
\cite{Wei}
distinguished between six various texture classes of sea ice dependent on the
crystal morphology, brine inclusions and  the genesis. The different ice
crystal growth depends on snow depostion, flooding, turbulent mixing, quiescent
growth rate or supercooling. Each condition determines the character of the
kinetics. Non-linear heat and salt dissipation for example leads to dendritic
growth (snowflakes) whereas one observes in sea ice mostly lamellar or
cellular structures rather than complete dendrits \cite{Ei}.  Hence, the
morphology of sea ice is one criterion for the choice of an appropriate
kinetics for the genesis of sea ice.  Therefore, in order to simulate different
structure sizes and textures, we can modify the dispersion relation by
varying the parameters $\alpha_1$, $\alpha_2$ and $D$ or by a modified
kinetics \cite{Tur}. The crucial point is the shape of the dispersion
function. If there are multiple different positive unstable regions for the
wavenumbers with positive real part of eigenvalues we could expect that
differently large channels evolve. For instance 
{\sc Worster et al}
  \cite{WW97} has presented a general theory for convection with mushy layers. The
two different minima of the neutral curve, determined by the linear stability
analysis, correspond to two different modes of convection, which affect the
kinetics and determines the size distribution of the brine
channels.  We note that the initial conditions are decisive for the
appearance of specific pattern \cite{Mu}. Hence one should investigate how
dislocations or antifreeze proteins influence the formation of the brine
channel distribution.




\appendix

\section{Dimensionless Quantities}

If we set $\tau=\frac{t}{t_0}$, $\xi=\frac{x}{x_0}$, $u=C_1\psi$ and $v=C_2
\rho$ we get with $\frac{\partial \psi}{\partial t}=\frac{\partial
  \tau}{\partial t}\frac{\partial \psi}{\partial
  \tau}=\frac{1}{t_0}\frac{\partial \psi}{\partial \tau}$ and $\frac{\partial \psi}{\partial x}=\frac{\partial
  \xi}{\partial x}\frac{\partial \psi}{\partial
  \xi}=\frac{1}{x_0}\frac{\partial \psi}{\partial \xi}$
\begin{equation}
\frac{\partial u}{\partial t}=C_1\frac{\partial \psi}{\partial t}=\frac{C_1}{t_0}\frac{\partial \psi}{\partial \tau}
\end{equation}
and 
\begin{equation}
\frac{\partial u}{\partial x}=C_1\frac{\partial \psi}{\partial
  x}=\frac{C_1}{x_0}\frac{\partial \psi}{\partial \xi}  .
\end{equation}
Because of $\frac{\partial^2 \psi}{\partial \xi^2}=\frac{\partial^2
  \psi}{\partial x^2} \left(\frac{\partial x}{\partial \xi} \right)^2
  +\frac{\partial \psi}{\partial x} \frac{\partial ^2 x}{\partial \xi^2}=x_0^2
  \frac{\partial^2 \psi}{\partial x^2}$ we obtain $\frac{\partial^2
  \psi}{\partial x^2}=\frac{1}{x^2_0}\frac{\partial^2 \psi}{\partial \xi^2}$
and consequently
\begin{equation}
\frac{\partial^2 u}{\partial x^2}=C_1\frac{\partial^2 \psi}{\partial
  x^2}=\frac{C_1}{x^2_0}\frac{\partial^2 \psi}{\partial \xi^2} .
\end{equation} 
Accordingly one has
\begin{eqnarray}
\frac{\partial v}{\partial t}&=&\frac{C_2}{t_0}\frac{\partial \rho}{\partial \tau}\nonumber\\
\frac{\partial^2 v}{\partial x^2}&=&\frac{C_2}{x^2_0}\frac{\partial^2 \rho}{\partial \xi^2} .
\end{eqnarray} 
From (\ref{f2}) and (\ref{f3}) follow the dimensionless equations
\begin{eqnarray}
\frac{\partial \psi}{\partial \tau}&=&
f[\psi,\rho]+D_1\frac{t_0}{x^2_0}\frac{\partial^2\psi}{\partial \xi^2}\\
\frac{\partial \rho}{\partial \tau}&=&
g[\psi,\rho]+D_2\frac{t_0}{x^2_0}\frac{\partial^2\rho}{\partial \xi^2}
\end{eqnarray} 
with
\begin{eqnarray}
f[\psi,\rho]&=& a_1 t_0 \psi - ct_0C^2_1\psi^3+dt_0C^4_1\psi^5+b_1t_0\frac{C_2}{C_1}\rho\nonumber\\
g[\psi,\rho]&=& -a_2 t_0 \rho - b_2t_0\frac{C_1}{C_2}\psi .
\end{eqnarray}
If we choose 
\begin{eqnarray}
&&ct_0C^2_1=1,\qquad
b_1t_0\frac{C_2}{C_1}=1,\nonumber\\
&&D_1\frac{t_0}{x^2_0}=1,\qquad
b_2t_0\frac{C_1}{C_2}=1
\end{eqnarray}
we obtain $C_1=\sqrt[4]{\frac{b_1 b_2}{c^2}}$,
$C_2=\sqrt[4]{\frac{b^3_2}{b_1c^2}}$, $t_0=\frac{1}{\sqrt{b_1 b_2}}$ and
$x_0=\sqrt[4]{\frac{D^2_1}{b_1 b_2}}$.

\section{Linear Stability Analysis}
Let $\bar \psi$ and $\bar \rho$ denote small displacements from the 
equili\-brium
values $\psi_0$ and $\rho_0$ and write  $\psi=\psi_0 + \bar \psi$ and
$\rho=\rho_0 + \bar \rho$. With respect to (\ref{f4}, \ref{f5}) we obtain
\begin{eqnarray}
  \bar \psi_\tau(\xi,\tau) & = &
  \alpha_1(\psi_0 + \bar \psi) - (\psi_0 + \bar \psi)^3 \nonumber \\ &+& \delta (\psi_0 +
  \bar \psi)^5 + \rho_0 +\bar \rho +
  \bar \psi_{\xi \xi}(\xi,\tau)
  \label{a1}\nonumber\\
  \bar \rho_\tau(\xi,\tau) & = &
  D \bar \rho_{\xi
  \xi}(\xi,\tau)-\alpha_2 (\rho_0 + \bar \rho) - (\psi_0 + \bar \psi) 
  \label{a2} .
\end{eqnarray}

If we consider only linear terms
\begin{eqnarray}
  \bar \psi_\tau(\xi,\tau) & = &
  \cdots + \alpha_1 \bar \psi - 3 \psi^2_0 \bar \psi \nonumber \\ &+& 5 \delta \psi^4_0 \bar \psi + \bar \rho + \cdots + \bar \psi_{\xi \xi}(\xi,\tau)\nonumber\\
  \bar \rho_\tau(\xi,\tau) & = &
  \cdots - \bar \psi -\alpha_2 \bar \rho + \cdots + D \bar \rho_{\xi
  \xi}(\xi,\tau)
\end{eqnarray}
we get 
\begin{eqnarray}
 \left( \begin{array}{c}
    \frac{\partial}{\partial{\tau}}\bar\psi\\\frac{\partial}{\partial{\tau}}\bar\rho  \end{array}
\right) &=& \underbrace{\left( \begin{array}{cc}\alpha_1-3\psi^2_0 +5\delta\psi^4_0 & 1\\-1 & -\alpha_2 \end{array}
\right)}_{\boldmath{J}_{(\psi=\psi_0, \rho=\rho_0)}} \left( \begin{array}{c} \bar\psi\\\bar\rho  \end{array} \right) \nonumber\\ &+& \left(
  \begin{array}{cc} \frac{\partial^2}{\partial{\xi^2}} & 0\\0 & D \frac{\partial^2}{\partial{\xi^2}} \end{array}
\right) \left( \begin{array}{c} \bar\psi\\\bar\rho  \end{array} \right) ,
\end{eqnarray}
where $\boldmath{J}_{(\psi=\psi_0, \rho=\rho_0)}$ is the Jacobian 
\begin{eqnarray}
\boldmath{J}_{(\psi=\psi_0, \rho=\rho_0)} &=& \left( \begin{array}{cc}f_\psi &
    f_\rho \nonumber\\g_\psi & g_\rho
  \end{array} \right)_{(\psi=\psi_0, \rho=\rho_0)}\\ &=&  \left( \begin{array}{cc}\alpha_1-3\psi^2_0+5\delta \psi^4_0 & 1\\-1 & -\alpha_2 \end{array}
\right) ,
\end{eqnarray}
which we can calculate also considering  (\ref{f5a}, \ref{f5aa}).

Using the Fourier ansatz $\bar \psi = \psi_0 exp(\lambda \tau
+i\kappa\xi)$ and $\bar \rho = \rho_0 exp(\lambda \tau +i\kappa\xi)$ in
(\ref{f7}) we find
\begin{eqnarray}
 \left( \begin{array}{c}
    \lambda \bar\psi_i\\ \lambda \bar\rho_i  \end{array}
\right) &=&   \left( \begin{array}{cc}\alpha_1-3\psi^2_0 +5\delta\psi^4_0 & 1\\-1 & -\alpha_2 \end{array}
\right) \left( \begin{array}{c} \bar\psi_i\\\bar\rho_i  \end{array} \right) \nonumber\\ &+& \left(
  \begin{array}{cc} -\kappa^2 & 0\\0 & -D \kappa^2 \end{array}
\right) \left( \begin{array}{c} \bar\psi_i\\\bar\rho_i  \end{array} \right) .
\end{eqnarray}
With $\psi_0=0$ and $\rho_0=0$ the eigenvalue equation
\begin{eqnarray}
 0 = \left[\!  
\left(\! 
\begin{array}{cc} \kappa^2\!-\!\alpha_1  & -1\\1 & \alpha_2\!+\! D\kappa^2 \end{array}\right) 
+ \left( \!\begin{array}{cc} \lambda & 0\\0 & \lambda \end{array}\right)
\right] 
\!\left( \begin{array}{c} \bar\psi_i\\\bar\rho_i  \end{array} \right)
\end{eqnarray}
follows with the characteristic equation
\begin{eqnarray}
 0&=&\left|\begin{array}{cc}\alpha_1 - \kappa^2 - \lambda & 1\\-1 & -\alpha_2-
 D\kappa^2 - \lambda \end{array} \right| \nonumber\\
&=& \lambda^2 +[\kappa^2(1+D)+\alpha_2-\alpha_1]\lambda \nonumber \\
&+& \underbrace{D\kappa^4+(\alpha_2-\alpha_1 D)\kappa^2-\alpha_1
 \alpha_2+1}_{h(\kappa^2)} .
\end{eqnarray}

\acknowledgments

The authors are deeply indebted  J\"urgen Weissenberger for the many
exposures of casts brine channels. Dr. L. D\"unkel and Dr. P. Augustin are thanked for their system theoretic discussion and Dr. U. Menzel at the Rudbeck laboratory, Uppsala, Sweden for his support.
This work was supported by DFG 
Priority Program 1157 via GE1202/06, 444BRA-113/57/0-1, the BMBF, the DAAD and by European 
ESF program NES. The financial support by the Brazilian Ministry of Science and Technology is acknowledged.

\bibliography{cauchy}
\end{document}